\documentclass[11pt,letterpaper]{article}
\usepackage{ijcnlp2017}
\usepackage{times}
\usepackage{latexsym}

\ijcnlpfinalcopy

\usepackage{helvet}
\usepackage{courier}
\usepackage{amsmath}
\usepackage{subcaption}
\usepackage{mathtools}
\usepackage{dblfloatfix} 
\usepackage{multirow}

\usepackage{algorithm}
\usepackage{algorithmic}

\usepackage{mathrsfs} 
\usepackage{graphicx}
\graphicspath{ {./images/} }

\title{Semantic Document Distance Measures and \\ Unsupervised Document Revision Detection}

\author{Xiaofeng Zhu, Diego Klabjan \\ 
  Northwestern University, USA \\
  {\tt xiaofengzhu2013@u.northwestern.edu}\\
  {\tt d-klabjan@northwestern.edu} \\\And
  Patrick N Bless \\
  Intel Corporation,\\ 
  Chandler, AZ, USA \\ 
  {\tt patrick.n.bless@intel.com}}

\date{}

\begin{document}

\maketitle
\begin{abstract}
In this paper, we model the document revision detection problem as a minimum cost branching problem that relies on computing document distances. Furthermore, we propose two new document distance measures, word vector-based Dynamic Time Warping (wDTW) and word vector-based Tree Edit Distance (wTED). Our revision detection system is designed for a large scale corpus and implemented in Apache Spark. We demonstrate that our system can more precisely detect revisions than state-of-the-art methods by utilizing the Wikipedia revision dumps \footnote{https://snap.stanford.edu/data/wiki-meta.html} and simulated data sets.
\end{abstract}

\section{Introduction}
\label{Introduction}

It is a common habit for people to keep several versions of documents, which creates duplicate data. A scholarly article is normally revised several times before being published. An academic paper may be listed on personal websites, digital conference libraries, Google Scholar, etc. In major corporations, a document typically goes through several revisions involving multiple editors and authors. Users would benefit from visualizing the entire history of a document. It is worthwhile to develop a system that is able to intelligently identify, manage and represent revisions. Given a collection of text documents, our study identifies revision relationships in a completely unsupervised way. For each document in a corpus we only use its content and the last modified timestamp. We assume that a document can be revised by many users, but that the documents are not merged together. We consider collaborative editing as revising documents one by one.

The two research problems that are most relevant to document revision detection are plagiarism detection and revision provenance. In a plagiarism detection system, every incoming document is compared with all registered non-plagiarized documents~\cite{si1997check,oberreuter2013text,hagen2015source,abdi2015pdlk}. The system returns true if an original copy is found in the database; otherwise, the system returns false and adds the document to the database. Thus, it is a 1-to-n problem. Revision provenance is a 1-to-1 problem as it keeps track of detailed updates of one document ~\cite{buneman2001and,zhang2013revision}. Real-world applications include GitHub, version control in Microsoft Word and Wikipedia version trees~\cite{sabel2007structuring}. In contrast, our system solves an n-to-n problem on a large scale. Our potential target data sources, such as the entire web or internal corpora in corporations, contain numerous original documents and their revisions. The aim is to find all revision document pairs within a reasonable time. 

Document revision detection, plagiarism detection and revision provenance all rely on comparing the content of two documents and assessing a distance/similarity score. The classic document similarity measure, especially for plagiarism detection, is fingerprinting~\cite{Hoad:2003:MIV:766441.766444,charikar2002similarity,schleimer2003winnowing, fujii2001japanese, manku2007detecting, manber1994finding}. Fixed-length fingerprints are created using hash functions to represent document features and are then used to measure document similarities. However, the main purpose of fingerprinting is to reduce computation instead of improving accuracy, and it cannot capture word semantics. Another widely used approach is computing the sentence-to-sentence Levenshtein distance and assigning an overall score for every document pair~\cite{Gustafson:2008:NHF:1486927.1486985}. Nevertheless, due to the large number of existing documents, as well as the large number of sentences in each document, the Levenshtein distance is not computation-friendly. Although alternatives such as the vector space model (VSM) can largely reduce the computation time, their effectiveness is low. More importantly, none of the above approaches can capture semantic meanings of words, which heavily limits the performances of these approaches. For instance, from a semantic perspective, ``\textit{I went to the bank}" is expected to be similar to ``\textit{I withdrew some money}" rather than ``\textit{I went hiking.}" Our document distance measures are inspired by the weaknesses of current document distance/similarity measures and recently proposed models for word representations such as word2vec ~\cite{mikolov2013distributed} and Paragraph Vector (PV) ~\cite{le2014distributed}. Replacing words with distributed vector embeddings makes it feasible to measure semantic distances using advanced algorithms, e.g., Dynamic Time Warping (DTW) ~\cite{sakurai2005ftw,muller2007dynamic,matuschek2008measuring} and Tree Edit Distance (TED) ~\cite{tai1979tree,zhang1989simple,klein1998computing,demaine2007optimal,pawlik2011rted,pawlik2014memory,pawlik2015efficient,pawlik2016tree}.  Although calculating text distance using DTW \cite{liu2007sentence}, TED \cite{sidorov2015computing} or Word Mover's Distance (WMV) \cite{kusner2015word} has been attempted in the past, these measures are not ideal for large-scale document distance calculation. The first two algorithms were designed for sentence distances instead of document distances. The third measure computes the distance of two documents by solving a transshipment problem between words in the two documents and uses word2vec embeddings to calculate semantic distances of words. The biggest limitation of WMV is its long computation time. We show in Section \ref{Results} that our wDTW and wTED measures yield more precise distance scores with much shorter running time than WMV.

We recast the problem of detecting document revisions as a network optimization problem (see Section \ref{Revision Network}) and consequently as a set of document distance problems (see Section \ref{wDTW and wTED Distance Measures}). We use trained word vectors to represent words, concatenate the word vectors to represent documents and combine word2vec with DTW or TED.  Meanwhile, in order to guarantee reasonable computation time in large data sets, we calculate document distances at the paragraph level with Apache Spark. A distance score is computed by feeding paragraph representations to DTW or TED. Our code and data are publicly available. \footnote{https://github.com/XiaofengZhu/wDTW-wTED}

The primary contributions of this work are as follows.
   \begin{itemize}
     \item We specify a model and algorithm to find the optimal document revision network from a large corpus.   
     \item We propose two algorithms, wDTW and wTED, for measuring semantic document distances based on distributed representations of words. The wDTW algorithm calculates document distances based on DTW by sequentially comparing any two paragraphs of two documents. The wTED method represents the section and subsection structures of a document in a tree with paragraphs being leaves. Both algorithms hinge on the distance between two paragraphs.
     \end{itemize}
The rest of this paper is organized in five parts. In Section 2, we clarify related terms and explain the methodology for document revision detection. In Section 3, we provide a brief background on existing document similarity measures and present our wDTW and wTED algorithms as well as the overall process flow. In Section 4, we demonstrate our revision detection results on Wikipedia revision dumps and six simulated data sets. Finally, in Section 5, we summarize some concluding remarks and discuss avenues for future work and improvements. 


\begin{figure*}
        \begin{subfigure}[b]{0.33\textwidth}
                \centering
                \includegraphics[width=.85\linewidth]{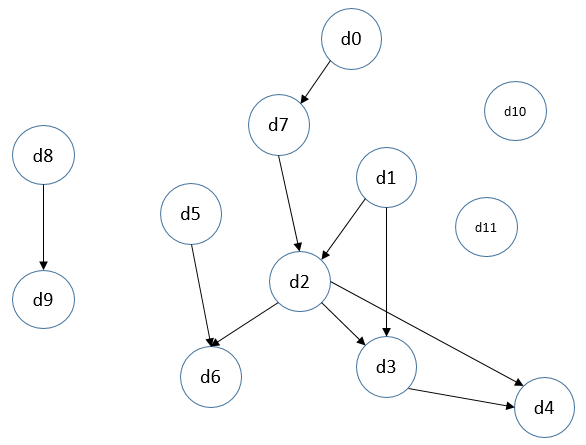}
                \caption{Revision network $N^0$}
                \label{fig:revision_network}
        \end{subfigure}%
        \begin{subfigure}[b]{0.33\textwidth}
                \centering
                \includegraphics[width=.85\linewidth]{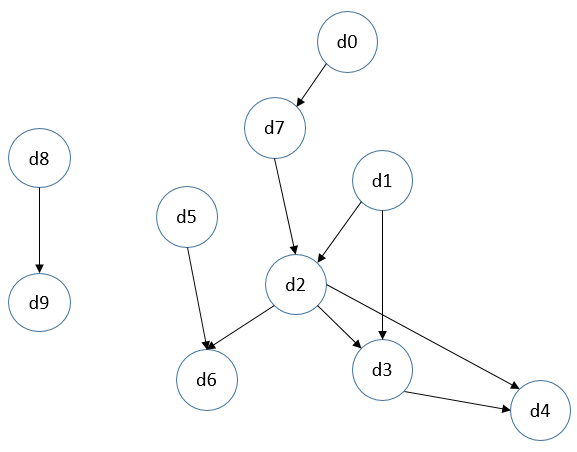}
                \caption{Cleaned revision network $N$}
                \label{fig:revision_network_cleaned}
        \end{subfigure}%
        \begin{subfigure}[b]{0.33\textwidth}
                \centering
                \includegraphics[width=.85\linewidth]{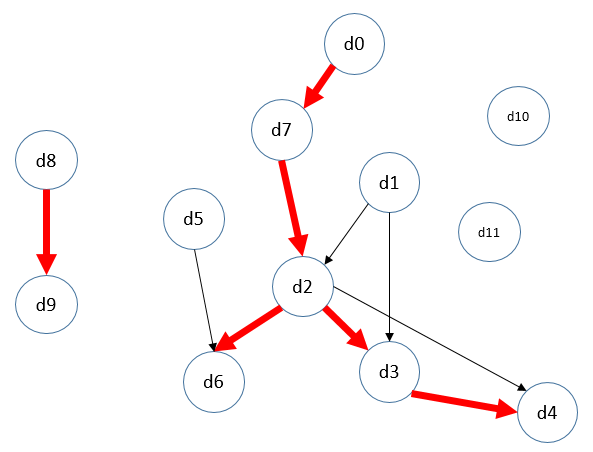}
                \caption{Possible solution $R$}
                \label{fig:revision_network_solution}
        \end{subfigure}%
        \caption{Revision network visualization}
\end{figure*}

\section{Revision Network}
\label{Revision Network}
The two requirements for a document $\bar{D}$ being a revision of another document $\tilde{D}$ are that $\bar{D}$ has been created later than $\tilde{D}$ and that the content of $\bar{D}$  is similar to (has been modified from) that of $\tilde{D}$. More specifically, given a corpus $ \mathscr{D}$, for any two documents $\bar{D}, \tilde{D} \in \mathscr{D}$, we want to find out the yes/no revision relationship of $\bar{D}$ and $\tilde{D}$, and then output all such revision pairs. 
%
%
%
%
%

We assume that each document has a creation date (the last modified timestamp) which is readily available from the meta data of the document. In this section we also assume that we have a $Dist$ method and a cut-off threshold $\tau$. We represent a corpus as network $N^0 = (V^0, A)$, for example Figure \ref{fig:revision_network}, in which a vertex corresponds to a document. There is an arc $a = (\bar{D}, \tilde{D})$ if and only if $Dist(\bar{D}, \tilde{D}) \leq \tau$ and the creation date of $\bar{D}$ is before the creation date of $\tilde{D}$. In other words, $\tilde{D}$ is a revision candidate for $\bar{D}$. By construction, $N^0$ is acyclic. For instance, $d_2$ is a revision candidate for $d_7$ and $d_1$. Note that we allow one document to be the original document of several revised documents.
 As we only need to focus on revision candidates, we reduce $N^0$ to $N = (V, A)$, shown in Figure \ref{fig:revision_network_cleaned}, by removing isolated vertices. We define the weight of an arc as the distance score between the two vertices. Recall the assumption that a document can be a revision of at most one document. In other words, documents cannot be merged. Due to this assumption, all revision pairs form a branching in $N$. (A branching is a subgraph where each vertex has an in-degree of at most 1.) The document revision problem is to find a minimum cost branching in $N$ (see Fig \ref{fig:revision_network_solution}).

The minimum branching problem was earlier solved by \citeauthor{edmonds1967optimum} \shortcite{edmonds1967optimum} and \citeauthor{velardi2013ontolearn} \shortcite{velardi2013ontolearn}. The details of his algorithm are as follows.
	\renewcommand{\labelitemi}{$\textendash –$}
	\begin{itemize}
		\item For each node select the smallest weighted incoming arc. This yields a subgraph.
		\item If cycles are present in the selected subgraph, then recursively find the replacing arc that has the minimum weight among previously non-selected arcs to eliminate cycles. 
	\end{itemize}

In our case, $N$ is acyclic and, therefore, the second step never occurs. For this reason, Algorithm \ref{algorithm:minimum_branching} solves the document revision problem.

\begin{algorithm}
  \caption{Find minimum branching $R$ for network $N = (V, A)$}

  \begin{algorithmic}[1]
	
	\STATE {\bfseries Input:} $N$
	\STATE $R = \emptyset$
			
		\FOR {every vertex $v \in V$} 
		\STATE Set $\delta(u)$ to correspond to all arcs with head $u$
		\STATE Select $a = (v, u) \in \delta(u)$ such that $Dist(v, u)$ is minimum 
		\STATE $R = R \cup {a}$
		\ENDFOR

	\STATE {\bfseries Output:} $R$
  \end{algorithmic}
  \label{algorithm:minimum_branching}
\end{algorithm}


The essential part of determining the minimum branching $R$ is extracting arcs with the lowest distance scores. This is equivalent to finding the most similar document from the revision candidates for every original document.

\section{Distance/similarity Measures}
\label{Document Distance measures based on Word2vec Embeddings}

In this section, we first introduce the classic VSM model, the word2vec model, DTW and TED.
We next demonstrate how to combine the above components to construct our semantic document distance measures: wDTW and wTED. We also discuss the implementation of our revision detection system.

\subsection{Background}
\label{Background}

\subsubsection{Vector Space Model}
\label{Vector Space Model}
VSM represents a set of documents as vectors of identifiers. The identifier of a word used in this work is the tf-idf weight.
We represent documents as tf-idf vectors, and thus the similarity of two documents can be described by the cosine distance between their vectors. VSM has low algorithm complexity but cannot represent the semantics of words since it is based on the bag-of-words assumption.
%

\subsubsection{Word2vec}
\label{Word2vec}
Word2vec produces semantic embeddings for words using a two-layer neural network. Specifically, word2vec relies on a skip-gram model that uses the current word to predict context words in a surrounding window to maximize the average log probability. Words with similar meanings tend to have similar embeddings.

\subsubsection{Dynamic Time Warping}
\label{Dynamic Time Warping}
DTW was developed originally for speech recognition in time series analysis and has been widely used to measure the distance between two sequences of vectors.

Given two sequences of feature vectors: $ A = a_1, a_2, ..., a_i, ..., a_m$ and $B = b_1, b_2, ..., b_j, ..., b_n$, DTW finds the optimal alignment for $A$ and $B$ by first constructing an $(m + 1) \times (n + 1)$ matrix in which the $(i, j)^{th}$ element is the alignment cost of $a_{1} ... a_{i}$ and $b_{1} ... b_{j}$, and then retrieving the path from one corner to the diagonal one through the matrix that has the minimal cumulative distance. This algorithm is described by the following formula.
\begin{equation*}
\begin{aligned}[t]
\label{equation:dtw}
DTW(i, j) &= Dist(a_i, b_j)+ min ( \\
&DTW(i-1, j), \ \ \ \ \ \ \ // insertion \\
&DTW(i, j-1), \ \ \ \ \ \ \ // deletion \\
&DTW(i-1, j-1)) \ // substitution
 \end{aligned}
\end{equation*}

\subsubsection{Tree Edit Distance}
\label{Tree Edit Distance}
TED was initially defined to calculate the minimal cost of node edit operations for transforming one labeled tree into another. The node edit operations are defined as follows.
   \begin{itemize}
     \item \textbf{Deletion} Delete a node and connect its children to its parent maintaining the order.   
     \item \textbf{Insertion} Insert a node between an existing node and a subsequence of consecutive children of this node.
     \item \textbf{Substitution} Rename the label of a node.
     \end{itemize}

Let $L_1$ and $L_2$ be two labeled trees, and $L_k[i]$ be the $i^{th}$ node in $L_k (k = 1, 2)$. $M$ corresponds to a mapping from $L_1$ to $L_2$. TED finds mapping $M$ with the minimal edit cost based on
\begin{equation*}
\begin{aligned}[t]
c(M) = min \{ &\sum_{(i, j)\in M} cost(L_1[i] \rightarrow L_2[j]) \\
&+  \sum_{i\in I} cost(L_1[i] \rightarrow \wedge) \\
&+  \sum_{j \in J} cost(\wedge \rightarrow L_2[j])
 \}
 \end{aligned}
\end{equation*}
where $L_1[i] \rightarrow L_2[j]$ means transferring $L_1[i]$ to $L_2[j]$ based on $M$, and $\wedge$ represents an empty node.

\subsection{Semantic Distance between Paragraphs}

According to the description of DTW in Section \ref{Dynamic Time Warping}, the distance between two documents can be calculated using DTW by replacing each element in the feature vectors $A$ and $B$ with a word vector. However, computing the DTW distance between two documents at the word level is basically as expensive as calculating the Levenshtein distance. Thus in this section we propose an improved algorithm that is more appropriate for document distance calculation.

In order to receive semantic representations for documents and maintain a reasonable algorithm complexity, we use word2vec to train word vectors and represent each paragraph as a sequence of vectors. Note that in both wDTW and wTED we take document titles and section titles as paragraphs. Although a more recently proposed model PV can directly train vector representations for short texts such as movie reviews ~\cite{le2014distributed}, our experiments in Section \ref{Results} show that PV is not appropriate for standard paragraphs in general documents. Therefore, we use word2vec in our work. Algorithm \ref{algorithm:dist_para} describes how we compute the distance between two paragraphs based on DTW and word vectors. The distance between one paragraph in a document and one paragraph in another document can be pre-calculated in parallel using Spark to provide faster computation for wDTW and wTED.

\begin{algorithm}
  \caption{DistPara}

  \begin{algorithmic}[h]
	\STATE {Replace the words in paragraphs $p$ and $q$ with word2vec embeddings: $\{v_i\}_{i=1}^{e}$ and $\{w_j\}_{j=1}^{f}$}
	\STATE {\bfseries Input:} $p = [v_1, .., v_e]$ and $q = [w_1, .., w_f]$	
	\STATE {Initialize the first row and the first column of $(e + 1) \times (f + 1)$ matrix $DTW_{para}$ $+ \infty$}
	\STATE {$DTW_{para}(0, 0) = 0$}
	\FOR {$i$ in range $(1, e + 1)$}
		\FOR {$j$ in range $(1, f + 1)$}	
			\STATE {$Dist(v_i, w_j) = || v_i - w_j||$}
			\STATE {calculate $DTW_{para}(i, j)$}
		\ENDFOR		
   	\ENDFOR
	\STATE {\bfseries Return:} $DTW_{para}(e, f)$
  \end{algorithmic}
  \label{algorithm:dist_para}
\end{algorithm}	

\section{wDTW and wTED Measures} 
\label{wDTW and wTED Distance Measures} 	
\subsection{Word Vector-based Dynamic Time Warping} 	
\label{wDTW}

\begin{figure}[!ht]

\centering
  \centering
  \includegraphics[width=0.5\textwidth]{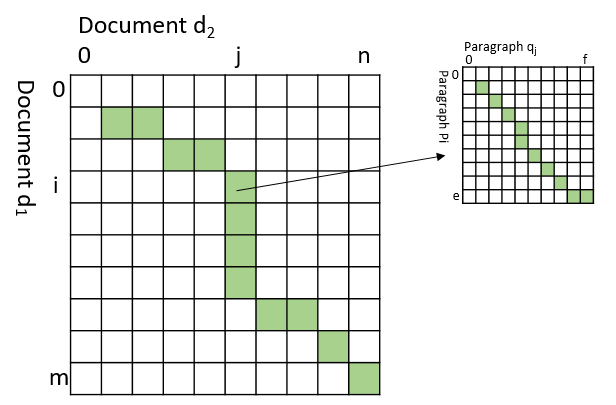}
  \caption{wDTW visualization}
  
\label{fig:wDTW}

\end{figure}

As a document can be considered as a sequence of paragraphs, wDTW returns the distance between two documents by applying another DTW on top of paragraphs. The cost function is exactly the DistPara distance of two paragraphs given in Algorithm \ref{algorithm:dist_para}. Algorithm \ref{algorithm:wDTW} and Figure \ref{fig:wDTW} describe our wDTW measure. wDTW observes semantic information from word vectors, which is fundamentally different from the word distance calculated from hierarchies among words in the algorithm proposed by ~\citeauthor{liu2007sentence} \shortcite{liu2007sentence}. The shortcomings of their work are that it is difficult to learn semantic taxonomy of all words and that their DTW algorithm can only be applied to sentences not documents.

\begin{algorithm}
  \caption{wDTW}

  \begin{algorithmic}[h]
  	\STATE {Represent documents $d_1$ and $d_2$ with vectors of paragraphs: $\{p_i\}_{i=1}^{m}$ and $\{q_j\}_{j=1}^{n}$}
	\STATE {\bfseries Input:} $d_1 = [p_1, .., p_m]$ and $d_2= [q_1, .., q_n]$	
	\STATE {Initialize the first row and the first column of $(m + 1) \times (n + 1)$ matrix $DTW_{doc}$ $+ \infty$}
	\STATE {$DTW_{doc}(0, 0) = 0$}
	\FOR {$i$ in range $(1, m + 1)$}
		\FOR {$j$ in range $(1, n + 1)$}	
			\STATE {$Dist(p_i, q_j) = $ DistPara$(p_i, q_j)$}
			\STATE {calculate $DTW_{doc}(i, j)$}
		\ENDFOR		
   	\ENDFOR
	\STATE {\bfseries Return:} $DTW_{doc}(m, n)$
  \end{algorithmic}
  \label{algorithm:wDTW}
\end{algorithm}	
	
\subsection{Word Vector-based Tree Edit Distance} 
\label{wTED}

\begin{figure}[h!]

\centering
  \centering
  \includegraphics[width= 0.4\textwidth]{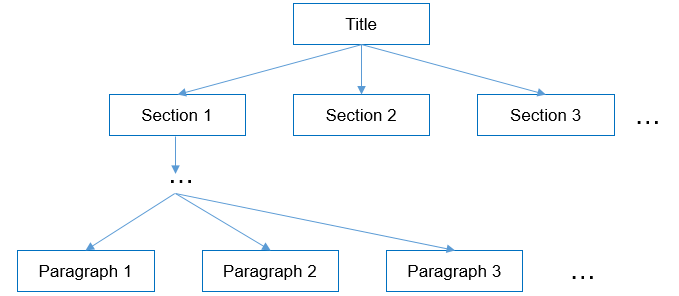}
  \caption{Document tree}
  
\label{fig:document_tree}

\end{figure}

TED is reasonable for measuring document distances as documents can be easily transformed to tree structures visualized in Figure \ref{fig:document_tree}. The document tree concept was originally proposed by \citeauthor{si1997check} \shortcite{si1997check}. A document can be viewed at multiple abstraction levels that include the document title, its sections, subsections, etc. Thus for each document we can build a tree-like structure with title $\rightarrow$ sections $\rightarrow$ subsections $\rightarrow$...$\rightarrow$ paragraphs being paths from the root to leaves. Child nodes are ordered from left to right as they appear in the document.
  
We represent labels in a document tree as the vector sequences of titles, sections, subsections and paragraphs with word2vec embeddings. wTED converts documents to tree structures and then uses DistPara distances. More formally, the distance between two nodes is computed as follows.
   \begin{itemize}
     \item The cost of substitution is the DistPara value of the two nodes.
     \item The cost of insertion is the DistPara value of an empty sequence and the label of the inserted node. This essentially means that the cost is the sum of the L2-norms of the word vectors in that node.
     \item The cost of deletion is the same as the cost of insertion.
   \end{itemize}

\begin{figure}[!ht]

\centering
  \centering
  \includegraphics[width=0.5\textwidth]{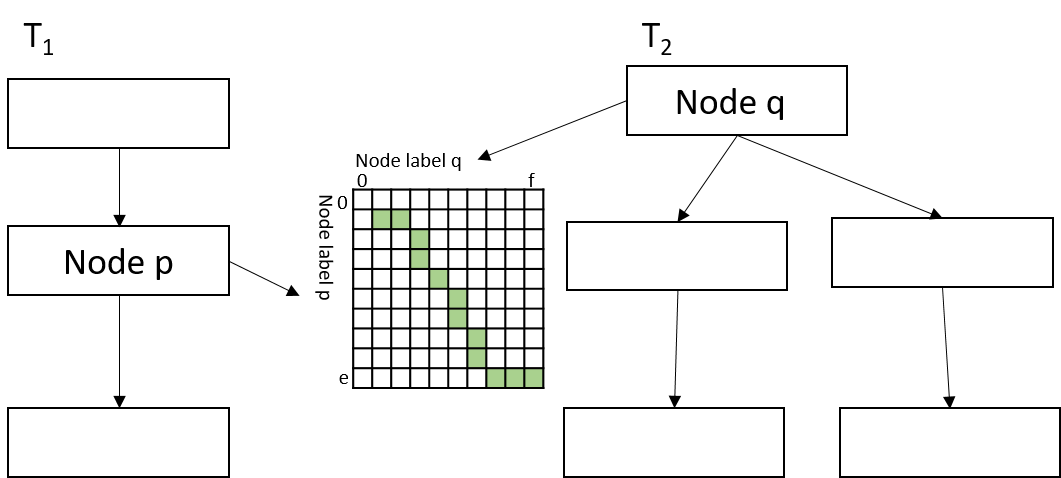}
  \caption{wTED visualization}
  
\label{fig:wTED}
\end{figure}

Compared to the algorithm proposed by \citeauthor{sidorov2015computing} \shortcite{sidorov2015computing}, wTED provides different edit cost functions and uses document tree structures instead of syntactic n-grams, and thus wTED yields more meaningful distance scores for long documents. Algorithm \ref{algorithm:wTED} and Figure \ref{fig:wTED} describe how we calculate the edit cost between two document trees.

\begin{algorithm}
  \caption{wTED}

  \begin{algorithmic}[1]
	\STATE {Convert documents $d_1$ and $d_2$ to trees $T_1$ and $T_2$}	
	\STATE {\bfseries Input:} $T_1$ and $T_2$

	\STATE {Initialize tree edit distance $c = + \infty$}	
	\FOR {node label $\ p \in T_1$}
		\FOR {node label $\ q \in T_2$}	
			\STATE {Update TED mapping cost $c$ using}
			\STATE {$cost(p \rightarrow q) = $ DistPara$(p, q)$}
			\STATE {$cost(p \rightarrow \wedge) = $ DistPara$(p, \wedge)$}
			\STATE {$cost(\wedge \rightarrow q) = $ DistPara$(\wedge, q)$}
		\ENDFOR		
   	\ENDFOR

	\STATE {\bfseries Return:} $c$
  \end{algorithmic}
  \label{algorithm:wTED}
\end{algorithm}

\subsection{Process Flow}
\label{Process Flow}
Our system is a boosting learner that is composed of four modules: weak filter, strong filter, revision network and optimal subnetwork. First of all, we sort all documents by timestamps and pair up documents so that we only compare each document with documents that have been created earlier. In the first module, we calculate the VSM similarity scores for all pairs and eliminate those with scores that are lower than an empirical threshold ($\tilde{\tau} = 0.5$). This is what we call the weak filter. After that, we apply the strong filter wDTW or wTED on the available pairs and filter out document pairs having distances higher than a threshold $\tau$. For VSM in Section \ref{Distance/Similarity Measures}, we directly filter out document pairs having similarity scores lower than a threshold $\tau$. The cut-off threshold estimation is explained in Section \ref{Estimating The Cut-off Threshold}. The remaining document pairs from the strong filter are then sent to the revision network module. In the end, we output the optimal revision pairs following the minimum branching strategy.

\subsection{Estimating the Cut-off Threshold}
\label{Estimating The Cut-off Threshold}

\begin{figure}[h!]
\centering
  \centering
  \includegraphics[width= 0.4\textwidth]{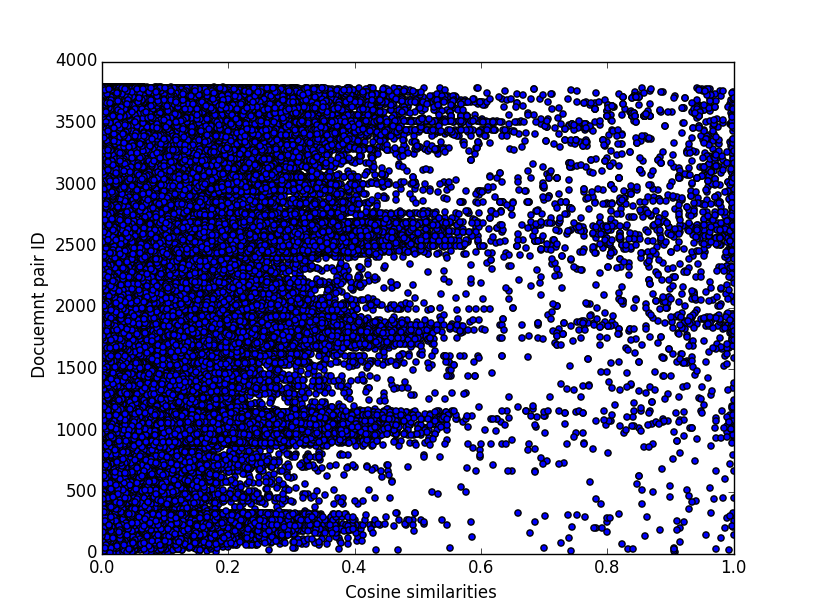}
  \caption{Setting $\tau$}
\label{fig:estimating tau}
\end{figure}

Hyperprameter $\tau$ is calibrated by calculating the absolute extreme based on an initial set of documents, i.e., all processed documents since the moment the system was put in use. Based on this set, we calculate all distance/similarity scores and create a histogram, see Figure \ref{fig:estimating tau}. 
The figure shows the correlation between the number of document pairs and the similarity scores in the training process of one simulated corpus using VSM. The optimal $\tau$ in this example is around 0.6 where the number of document pairs noticeably drops.

As the system continues running, new documents become available and $\tau$ can be periodically updated by using the same method.

\section{Numerical Experiments}
\label{Numerical Experiments}
This section reports the results of the experiments conducted on two data sets for evaluating the performances of wDTW and wTED against other baseline methods.

\subsection{Distance/Similarity Measures}
\label{Distance/Similarity Measures}
We denote the following distance/similarity measures.
   \begin{itemize}
		\item  wDTW: Our semantic distance measure explained in Section \ref{wDTW}. 
		
		\item  wTED: Our semantic distance measure explained in Section \ref{wTED}.				 

		\item WMD: The Word Mover's Distance introduced in Section \ref{Introduction}. WMD adapts the earth mover's distance to the space of documents.

		\item  VSM: The similarity measure introduced in Section \ref{Vector Space Model}. 
				
		\item PV-DTW: PV-DTW is the same as Algorithm \ref{algorithm:wDTW} except that the distance between two paragraphs is not based on Algorithm \ref{algorithm:dist_para} but rather computed as $||PV(p_1) - PV(p_2)||$ where $PV(p)$ is the PV embedding of paragraph $p$. 

		\item PV-TED: PV-TED is the same as Algorithm \ref{algorithm:wTED} except that the distance between two paragraphs is not based on Algorithm \ref{algorithm:dist_para} but rather computed as $||PV(p_1) - PV(p_2)||$.

   \end{itemize}
Our experiments were conducted on an Apache Spark cluster with 32 cores and 320 GB total memory. We implemented wDTW, wTED, WMD, VSM, PV-DTW and PV-TED in Java Spark. The paragraph vectors for PV-DTW and PV-TED were trained by gensim. \footnote{https://radimrehurek.com/gensim/models/doc2vec.html} 

\subsection{Data Sets}
In this section, we introduce the two data sets we used for our revision detection experiments: Wikipedia revision dumps and a document revision data set generated by a computer simulation. The two data sets differ in that the Wikipedia revision dumps only contain linear revision chains, while the simulated data sets also contains tree-structured revision chains, which can be very common in real-world data. 
%
%
%

\subsubsection{Wikipedia Revision Dumps}
The Wikipedia revision dumps that were previously introduced by Leskovec et al. ~\shortcite{leskovec2010governance} contain eight GB (compressed size) revision edits with meta data. 

We pre-processed the Wikipedia revision dumps using the JWPL Revision Machine ~\cite{ferschke-zesch-gurevych:2011:Demos} and produced a data set that contains 62,234 documents with 46,354 revisions. As we noticed that short documents just contributed to noise (graffiti) in the data, we eliminated documents that have fewer than three paragraphs and fewer than 300 words. We removed empty lines in the documents and trained word2vec embeddings on the entire corpus. We used the documents occurring in the first $80\%$ of the revision period for $\tau$ calibration, and the remaining documents for test.

\begin{figure}[h!]

\centering
  \centering
  \includegraphics[width= 0.5\textwidth]{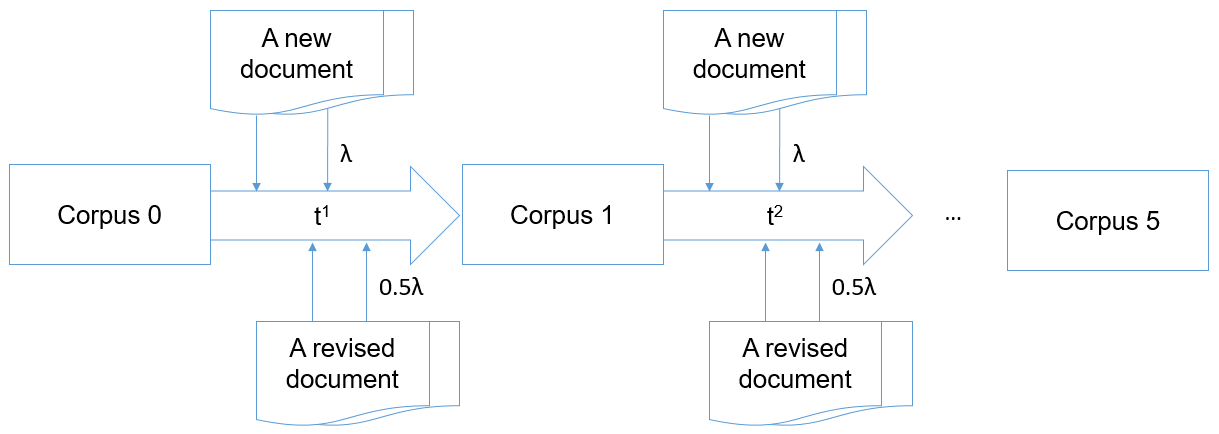}
  \caption{Corpora simulation}
  
\label{fig:revision_simulation}
\end{figure}

\begin{figure*}[!b]
        \begin{subfigure}[b]{0.33\textwidth}
                \centering
                \includegraphics[width=\linewidth]{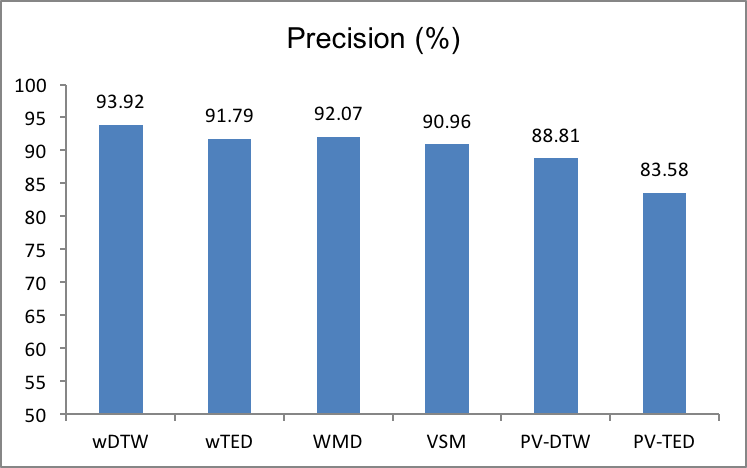}
                \caption{Precision}
                \label{fig:wiki_precision}
        \end{subfigure}%
        \begin{subfigure}[b]{0.33\textwidth}
                \centering
                \includegraphics[width=\linewidth]{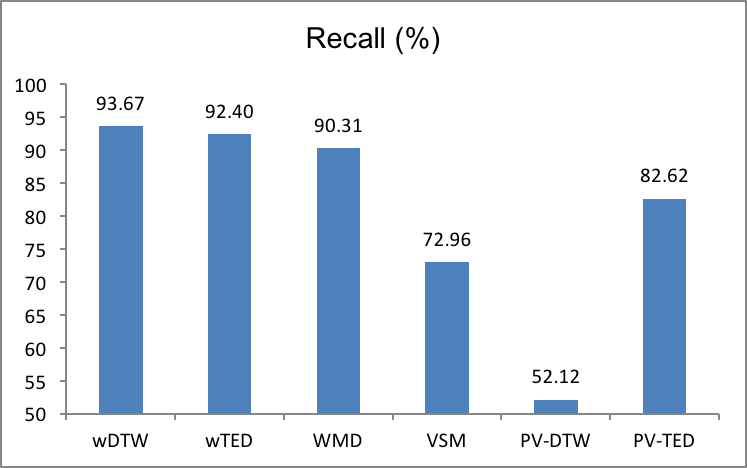}
                \caption{Recall}
                \label{fig:wiki_recall}
        \end{subfigure}%
        \begin{subfigure}[b]{0.33\textwidth}
                \centering
                \includegraphics[width=\linewidth]{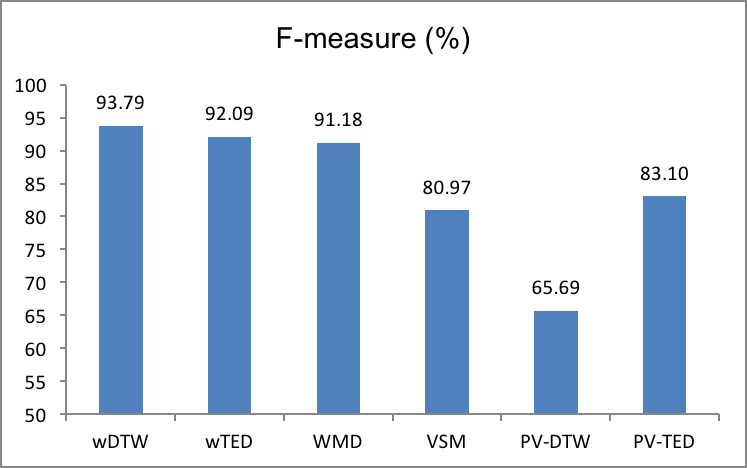}
                \caption{F-measure}
                \label{fig:wiki_fmeasure}
        \end{subfigure}%
        
        \caption{Precision, recall and F-measure on the Wikipedia revision dumps}
        \label{fig:result_wiki}
\end{figure*}

\subsubsection{Simulated Data Sets}

The generation process of the simulated data sets is designed to mimic the real world. Users open some existing documents in a file system, make some changes (e.g. \textit{addition}, \textit{deletion} or \textit{replacement}), and save them as separate documents. These documents become revisions of the original documents. We started from an initial corpus that did not have revisions, and kept adding new documents and revising existing documents. Similar to a file system, at any moment new documents could be added and/or some of the current documents could be revised. The revision operations we used were \textit{deletion}, \textit{addition} and \textit{replacement} of words, sentences, paragraphs, section names and document titles. The \textit{addition} of words, ..., section names, and new documents were pulled from the Wikipedia abstracts. This corpus generation process had five time periods $\{t^1, t^2, ...,  t^5\}$. Figure \ref{fig:revision_simulation} illustrates this simulation. We set a Poisson distribution with rate $\lambda = 550$ (the number of documents in the initial corpus) to control the number of new documents added in each time period, and a Poisson distribution with rate $0.5 \lambda$ to control the number of documents revised in each time period. 


We generated six data sets using different random seeds, and each data set contained six corpora (Corpus 0 - 5). Table \ref{table:simulated} summarizes the first data set. In each data set, we name the initial corpus Corpus 0, and define $T_0$ as the timestamp when we started this simulation process.  We set $T_{j} = T_{j-1} + t^{j}$, $j \in [1, 5]$. 
Corpus $j$ corresponds to documents generated before timestamp $T_j$. We extracted document revisions from Corpus $k \in [2, 5]$ and compared the revisions generated in (Corpus $k$ - Corpus $(k - 1)$) with the ground truths in Table \ref{table:simulated}. Hence, we ran four experiments on this data set in total. In every experiment, $\tau^k$ is calibrated based on Corpus $(k - 1)$. For instance, the training set of the first experiment was Corpus 1. We trained $\tau^1$ from Corpus 1. We extracted all revisions in Corpus 2, and compared revisions generated in the test set (Corpus 2 - Corpus 1) with the ground truth: 258 revised documents. 
The word2vec model shared in the four experiments was trained on Corpus $5$.

\begin{table}[h!]
\centering
\small
\caption{A simulated data set}
\label{table:simulated}
\begin{tabular}{rrrr}
\multicolumn{1}{l}{\multirow{2}{*}{Corpus}} & \multicolumn{1}{l}{Number of} & \multicolumn{1}{l}{Number of}     & \multicolumn{1}{l}{Number of}      \\
\multicolumn{1}{l}{}                        & \multicolumn{1}{l}{documents} & \multicolumn{1}{l}{new documents} & \multicolumn{1}{l}{revision pairs} \\
0	&550					&0				&0 \\
1	&1347				&542				&255 \\
2	&2125				&520				&258 \\
3	&2912				&528				&259 \\
4	&3777				&580				&285 \\
5	&4582				&547				&258 \\
\end{tabular}
\end{table}

\begin{figure*}[h!]
        \begin{subfigure}[b]{0.33\textwidth}
                \centering
                \includegraphics[width=\linewidth]{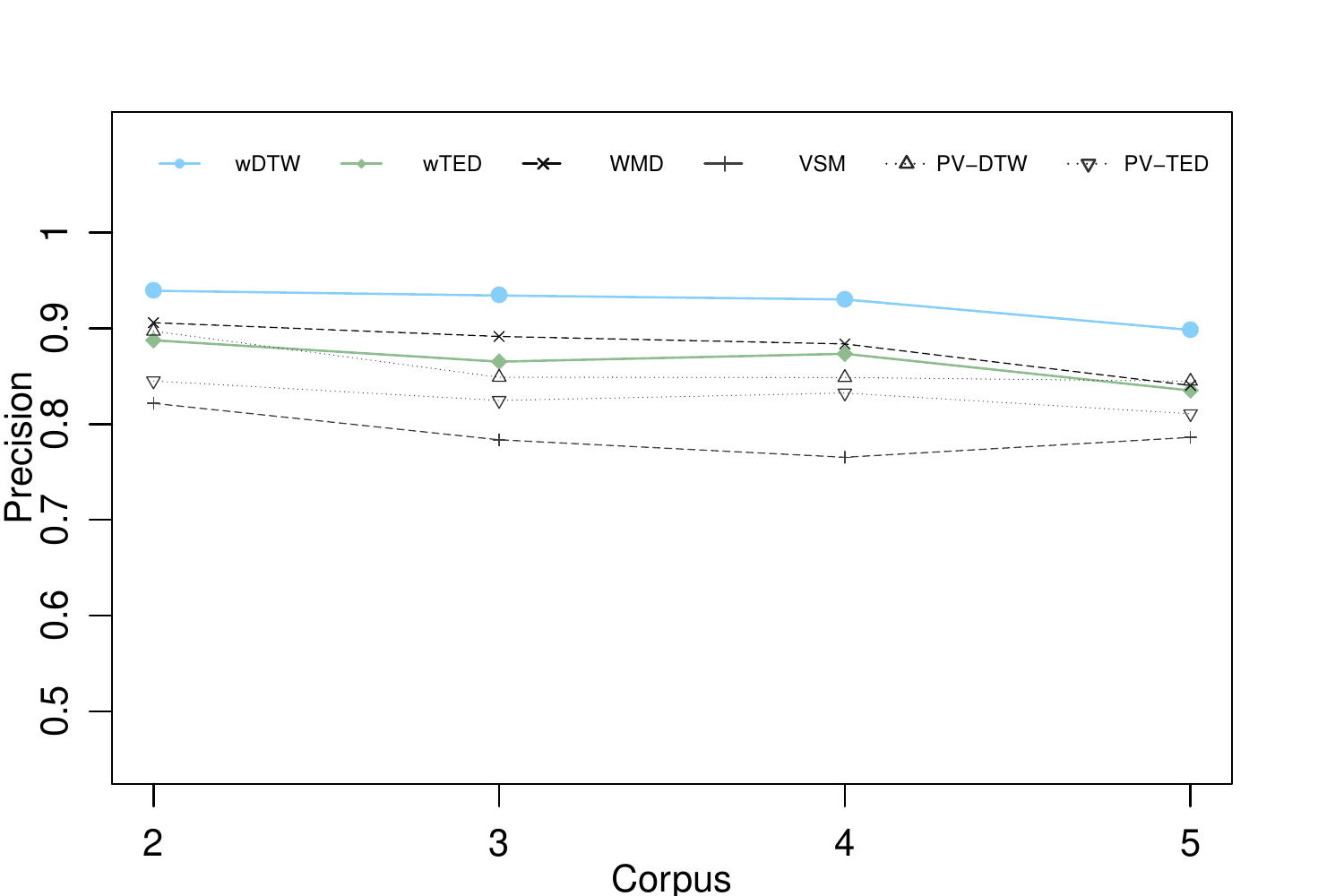}
                \caption{Precision}
                \label{fig:document_revision_precision}
        \end{subfigure}%
        \begin{subfigure}[b]{0.33\textwidth}
                \centering
                \includegraphics[width=\linewidth]{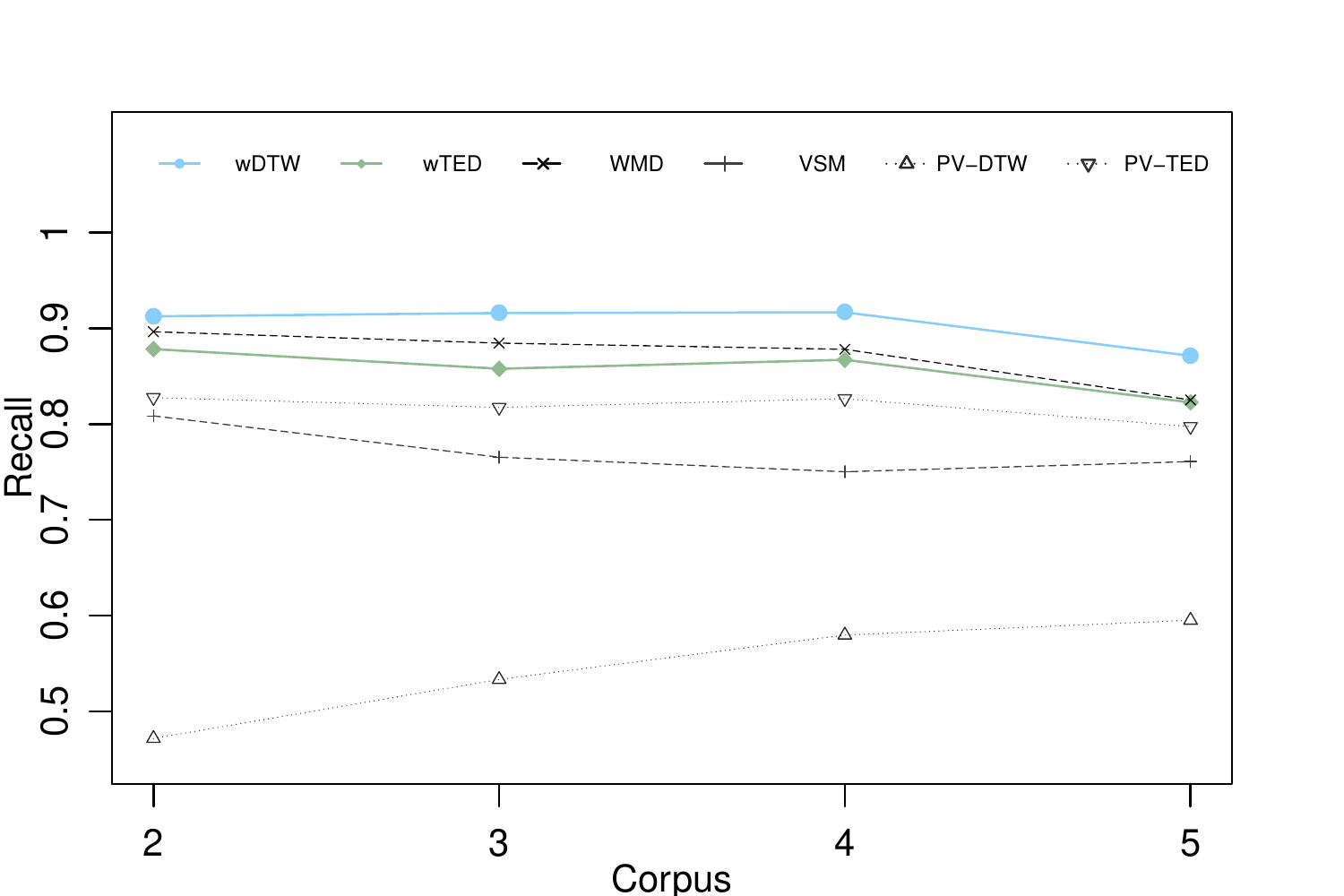}
                \caption{Recall}
                \label{fig:document_revision_recall}
        \end{subfigure}%
        \begin{subfigure}[b]{0.33\textwidth}
                \centering
                \includegraphics[width=\linewidth]{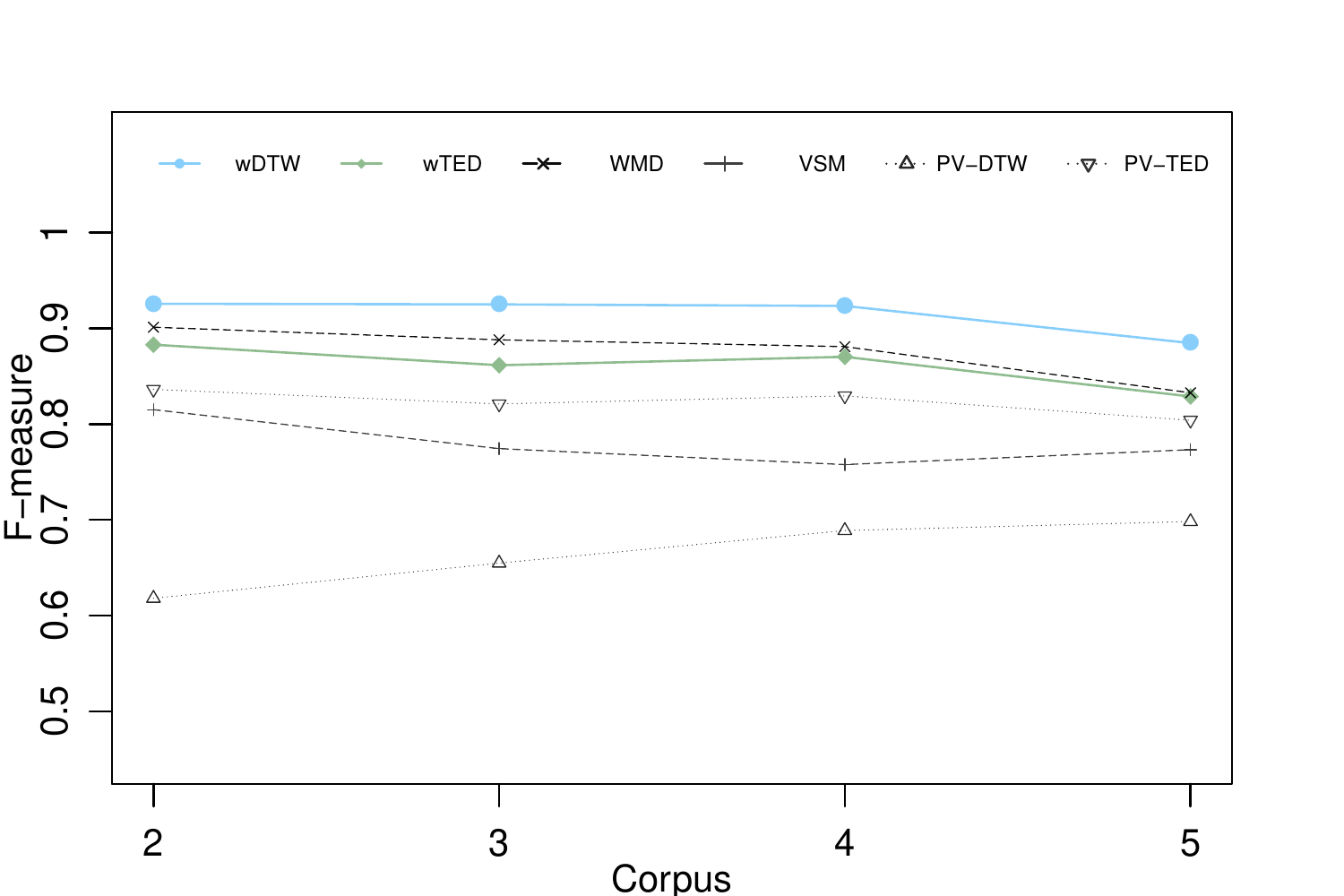}
                \caption{F-measure}
                \label{fig:document_revision_fmeasure}
        \end{subfigure}%
        
        \caption{Average precision, recall and F-measure on the simulated data sets}
        \label{fig:result_simulated}
\end{figure*}

\begin{table*}[h!]
\begin{center}
\caption{Running time of VSM, PV-TED, PV-DTW, wTED, wDTW and WMD}
\scalebox{0.9}{
\begin{tabular}{l|rrrrrr}
 &{VSM} &{PV-TED} &{PV-DTW}  &{wTED} &{wDTW} &{WMD} \\
Wikipedia revision dumps &1h 38min &3h 2min &3h 18min &5h 13min &13h 27min &515h 9min \\
corpus 2 &2 min &3 min &3 min &7 min &8 min	&8 h 53 min \\		
corpus 3 &3 min &4 min &5 min &9 min &11 min &12 h 45 min \\			                            
corpus 4 &4 min &6 min &6 min &11 min &12 min &14 h 34 min \\
corpus 5 &7 min &9 min &9 min &14 min &16 min &17 h 31 min  \\ 
\label{table: running time}		
\end{tabular}
}
\end{center}
\end{table*}

\subsection{Results}
\label{Results}
We use precision, recall and F-measure to evaluate the detected revisions. A true positive case is a correctly identified revision. A false positive case is an incorrectly identified revision. A false negative case is a missed revision record. 

We illustrate the performances of wDTW, wTED, WMD, VSM, PV-DTW and PV-TED on the Wikipedia revision dumps in Figure \ref{fig:result_wiki}. wDTW and wTED have the highest F-measure scores compared to the rest of four measures, and wDTW also have the highest precision and recall scores. Figure \ref{fig:result_simulated} shows the average evaluation results on the simulated data sets. From left to right, the corpus size increases and the revision chains become longer, thus it becomes more challenging to detect document revisions. 
Overall, wDTW consistently performs the best. WMD is slightly better than wTED. In particular, when the corpus size increases, the performances of WMD, VSM, PV-DTW and PV-TED drop faster than wDTW and wTED. Because the revision operations were randomly selected in each corpus, it is possible that there are non-monotone points in the series. 

wDTW and wTED perform better than WMD especially when the corpus is large, because they use dynamic programming to find the global optimal alignment for documents. In contrast, WMD relies on a greedy algorithm that sums up the minimal cost for every word. wDTW and wTED perform better than PV-DTW and PV-TED, which indicates that our DistPara distance in Algorithm \ref{algorithm:dist_para} is more accurate than the Euclidian distance between paragraph vectors trained by PV. 

We show in Table \ref{table: running time} the average running time of the six distance/similarity measures. In all the experiments, VSM is the fastest, wTED is faster than wDTW, and WMD is the slowest. Running WMD is extremely expensive because WMD needs to solve an $x^2$ sequential transshipment problem for every two documents where $x$ is the average number of words in a document. In contrast, by splitting this heavy computation into several smaller problems (finding the distance between any two paragraphs), which can be run in parallel, wDTW and wTED scale much better. 

Combining Figure \ref{fig:result_wiki}, Figure \ref{fig:result_simulated} and Table \ref{table: running time} we conclude that wDTW yields the most accurate results using marginally more time than VSM, PV-TED and PV-DTW, but much less running time than WMD. wTED returns satisfactory results using shorter time than wDTW.

\section{Conclusion}

This paper has explored how DTW and TED can be extended with word2vec to construct semantic document distance measures: wDTW and wTED. By representing paragraphs with concatenations of word vectors, wDTW and wTED are able to capture the semantics of the words and thus give more accurate distance scores. In order to detect revisions, we have used minimum branching on an appropriately developed network with document distance scores serving as arc weights. We have also assessed the efficiency of the method of retrieving an optimal revision subnetwork by finding the minimum branching.

Furthermore, we have compared wDTW and wTED with several distance measures for revision detection tasks. Our results demonstrate the effectiveness and robustness of wDTW and wTED in the Wikipedia revision dumps and our simulated data sets. In order to reduce the computation time, we have computed document distances at the paragraph level and implemented a boosting learning system using Apache Spark. Although we have demonstrated the superiority of our semantic measures only in the revision detection experiments, wDTW and wTED can also be used as semantic distance measures in many clustering, classification tasks.

Our revision detection system can be enhanced with richer features such as author information and writing styles, and exact changes in revision pairs. Another interesting aspect we would like to explore in the future is reducing the complexities of calculating the distance between two paragraphs.


\section*{Acknowledgments}
This work was supported in part by Intel Corporation, Semiconductor Research Corporation (SRC).  

\bibliography{DRD_IJCNLP2017}
\bibliographystyle{ijcnlp2017}

\end{document}